\documentclass[pdflatex,sn-mathphys-num]{sn-jnl}

\usepackage{graphicx}%
\usepackage{multirow}%
\usepackage{amsmath,amssymb,amsfonts}%
\usepackage{amsthm}%
\usepackage{mathrsfs}%
\usepackage[title]{appendix}%
\usepackage{xcolor}%
\usepackage{textcomp}%
\usepackage{manyfoot}%
\usepackage{booktabs}%
\usepackage{algorithm}%
\usepackage{algorithmicx}%
\usepackage{algpseudocode}%
\usepackage{listings}%
\usepackage{lineno}

\theoremstyle{thmstyleone}%
\theoremstyle{thmstyletwo}%

\theoremstyle{thmstylethree}%

\begin{document}

\title[Article Title]{Field-effect detected magnetic resonance of NV centers in diamond based on all-carbon Schottky contacts }

\author[1]{\fnm{Xuan Phuc} \sur{LE}}
\author[1]{\fnm{Ludovic} \sur{MAYER}}
\author[1]{\fnm{Simone} \sur{MAGALETTI}}
\author[2]{\fnm{Martin} \sur{SCHMIDT}}
\author[2]{\fnm{Jean-François} \sur{ROCH}}
\author[1]{\fnm{Thierry} \sur{DEBUISSCHERT}}
\affil[1]{Thales Research and Technology, 1 Avenue Augustin Fresnel, 91767 Palaiseau Cedex, France}
\affil[2]{Université Paris-Saclay, CNRS, ENS Paris-Saclay, CentraleSupelec, LuMIn, 91190 Gif-sur-Yvette, France}


    \abstract{The nitrogen vacancy (NV) center is a defect in diamond whose spin state can be read optically by exploiting its photoluminescence or electrically by exploiting its charge generation rate under illumination, both of which being spin-dependent. The latter method offers numerous opportunities in terms of integration and performance compared to conventional optical reading. Here, we investigate the physical properties of a graphitic-diamond-graphitic structure under illumination. We show how, for a type IIa diamond material, electron-hole pairs generated by an ensemble of NV centers lead to a p-type material upon illumination, making this all-carbon structure equivalent to two back-to-back Schottky diodes. We analyze how the reverse current flowing upon illumination changes as a function of bias voltage and radiofrequency-induced excitation of the NV ensemble spin resonances. Furthermore, we demonstrate how an additional field effect arising from the illumination scheme affects the reverse current, resulting in a photoelectrical signal that can exceed the optical signal under the same illumination conditions.}




\maketitle

\section{Introduction}\label{sec1}

The nitrogen-vacancy (NV) center in diamond is a color center that has shown extremely promising for quantum technology applications such as magnetometry \cite{rondin2014magnetometry, maze2008nanoscale}, electrometry \cite{chen2017high}, thermometry \cite{fukami2019all, maze2008nanoscale}, radiofrequency sensing \cite{magaletti2022quantum}, quantum networks \cite{nemoto2016photonic, ruf2021quantum, pompili2021realization}, and quantum computing \cite{childress2013diamond, nguyen2023software}. Its strength lies in the optical polarization, coherent manipulation, and readout of its electron spin state at room temperature on millisecond timescales. The NV center spin state readout relies generally on the collection of its spin-dependent photoluminescence. However, due to the high refractive index of diamond, only a limited amount of emitted photons can be detected. An alternative to overcome the limitation of the optical spin readout is the electrical spin readout \cite{bourgeois2021fundaments}. It relies on a photoconductive detection scheme where the spin-dependent charge generation rate from the NV centers upon illumination is measured by monitoring the photocurrent between ohmic contacts on the diamond surface \cite{bourgeois2020photoelectric}. The ohmic nature of the contact allows collecting a photocurrent linearly dependent on the number of charges generated in the diamond by NV centers and other defects. We previously demonstrated in a proof-of-principle experiment that the photoconductive detection of the NV centers spin resonance can be implemented with graphitic electrodes within the diamond sample, then opening the way to all-carbon sensing devices \cite{villaret2023efficient}.

Here we investigate the physical properties under illumination of a graphitic-diamond-graphitic structure similar to the one of reference \cite{villaret2023efficient}, and we demonstrate that it is equivalent to two Schottky diodes in a back-to-back configuration. We jointly use both optical and electrical detection of the NV centers spin resonance to identify the conduction state of the diodes and we show that the electrical signal coming from NV centers originates from the reverse-biased diode. We observe how the photogenerated carriers from NV centers produce a depletion region (DR) near the Schottky contact that stabilizes NV centers in their negative charge state due to band bending, indicating that our diamond sample behaves as a p-type semiconductor upon green illumination. We analyze how the reverse current flowing in the graphitic-diamond-graphitic structure changes with the bias voltage, and with the excitation of the spin resonances of the NV centers induced by a radiofrequency field. Furthermore, we demonstrate that an additional field effect, arising from the lateral illumination scheme used in the experiment, impacts the reverse current, resulting in an electrical contrast that can exceed the optical contrast under the same illumination conditions.

\section{Spin dependent charge states conversion of NV center in diamond}\label{sec2}

The nitrogen-vacancy center in diamond consists of a substitutional nitrogen atom associated to a carbon vacancy in two adjacent positions in the diamond lattice. It exists in different charge states. In its neutral charge state (NV$^{0}$), it has a single unpaired electron with spin one-half (S = 1/2) \cite{gali2009theory}. The negative charge state (NV$^{-}$) has two unpaired electrons, leading to spin triplet (S = 1) and spin singlet (S = 0) states \cite{Anishchik_2015}. The energy levels corresponding to these two charge states are depicted in Figure \ref{energyband}-a. NV$^{0}$ has a zero phonon line (ZPL) at 575 nm, while NV$^{-}$ has a ZPL at 637 nm \cite{rondin2010surface}. Under green optical pumping (561 nm in our situation), the NV center cycles continuously between its two charge states. The conversion of NV$^{-}$ to NV$^{0}$ through a two-photon absorption process (photoionization) generates a free electron in the diamond conduction band. Similarly, the two-photon backconversion process from NV$^{0}$ to NV$^{-}$ leaves a free hole in the diamond valence band. Consequently, the NV$^-\rightarrow$ NV$^0\rightarrow$ NV$^-$ cycle leaves a free electron-hole pair which results in a photocurrent (PC) when a voltage is applied \cite{bourgeois2015photoelectric}.    
Concerning NV$^{-}$, the ground state (G.S.) and the excited state (E.S.) are spin triplets, and their spin sublevels are labeled by the spin projection quantum number ($\mathrm{m_s}$) along the nitrogen-vacancy quantization axis. The zero-field splitting (ZFS) between the $\mathrm{\left| {{m_S} = 0} \right\rangle}$ state and the two degenerate $\mathrm{\left| {{m_S} = \pm 1} \right\rangle}$ ground state spin sublevels is 2.87 GHz \cite{kim2019cmos}. All sublevels are equally populated at room temperature. Two decay paths exist from the excited state to the ground state: first, a spin-conserving radiative transition, which is responsible of the photoluminescence (PL) emission; second, a non-radiative intersystem crossing through a metastable singlet state (M.S.). The intersystem crossing transition rate is higher from the $\mathrm{\left|{m_S} = \pm 1\right \rangle}$ spin sublevels than the $\mathrm{\left|{m_S}=0\right\rangle}$ spin sublevel of the excited state, resulting in the optical polarization of the NV$^-$ center in the $\mathrm{\left|{m_S} = 0\right\rangle}$ ground state and a decrease in both the PL emission intensity and the charge conversion rate \cite{tetienne2012magnetic}. As a consequence, under optical pumping, the spin magnetic resonance of NV$^{-}$ can be both optically (ODMR, \cite{manson2006nitrogen}) and electrically (PDMR, \cite{lin2008one,PhysRevLett.106.157601,sola2019electron}) detected by monitoring respectively the PL and PC intensity. 

\renewcommand{\figurename}{\textbf{Fig}}
\makeatletter
\renewcommand{\thefigure}{\textbf{\arabic{figure}}}
\makeatother
\begin{figure*}[h!]
    \centering
    \includegraphics[width=\textwidth]{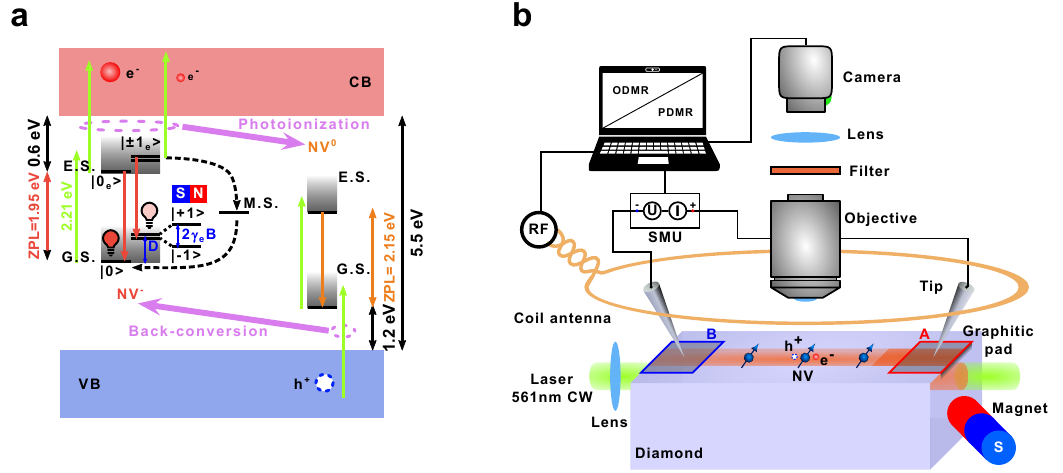}
    \caption{\textbf{NV center photophysical properties and the sketch of the experiment exploiting both optical and electrical readout.} \textbf{a}, Representation of NV$^0$ and NV$^-$ energy levels inside the 5.5 eV diamond band gap and their charge switching mechanism. The zero-field splitting of NV$^-$ is D = 2.87 GHz. The green, orange, red, and dashed arrows represent respectively the excitation process, the radiative relaxation of NV$^0$, the radiative relaxation of NV$^-$ and the non-radiative relaxation of NV$^-$. The two-photon ionization of NV$^-$ and the two-photon back-conversion of NV$^0$ generate a free electron-hole pair, in the conduction and valence band, respectively. Applying a microwave field allows both the electrical and optical spin-readout. \textbf{b}, Sketch of the experiment. Two graphitic electrodes, of surface 50$\times$50 \textmu m$^2$, are separated by 200 µm of distance and denoted A (in red) and B (in blue). The pale green and red cylinders materialize the 561 nm laser and the photoluminescence signal generated from NV centers in the bulk diamond. This illumination connects the two electrodes. Two probes are used to apply a bias voltage and detect the photocurrent for the PDMR imaging. A loop antenna covers the entire system to generate the RF signal and drive the NV$^-$ spin transitions. A microscope objective and a camera are used to image the photoluminescence from the NV centers signal and to perform ODMR measurements.} 
    \label{energyband}
    
\end{figure*}

\section{Experimental results}\label{sec4}

\begin{figure*}[h]
    \centering
    \includegraphics[width=\textwidth]{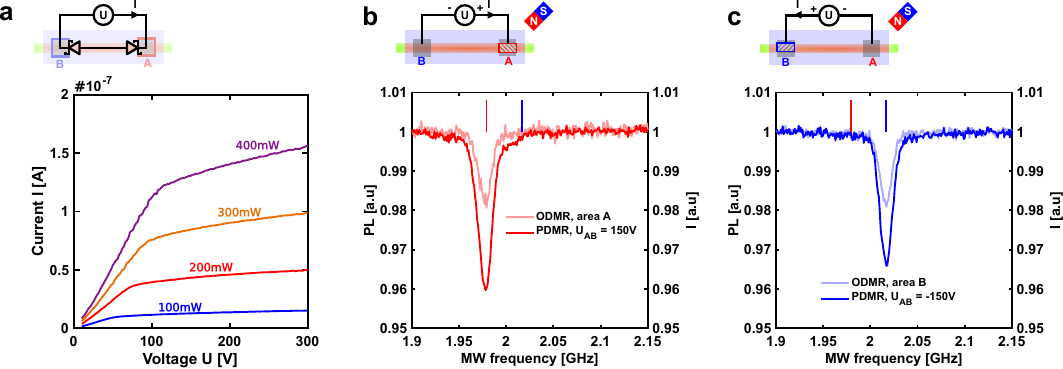}
    \caption{\textbf{Electrical and optical characterization of the graphitic-diamond-graphitic structure and identification of the conduction state of back-to-back Schottky contacts using ODMR/PDMR measurements.} \textbf{a}, I-U characteristics of the graphitic-diamond-graphitic junction under illumination with different optical power and the equivalent circuit with two back-to-back Schottky diodes. \textbf{b} and \textbf{c}, Identification of the contact polarity and the conduction state of the two back-to-back Schottky contacts exploiting PDMR and ODMR signals under a magnetic field gradient. The dark red (or blue) curve represents the normalized PC signal at electrode A (or B). The light red (or blue) curve represents the normalized PL signal at electrode A (or B).} 
    \label{exp1}
\end{figure*}

PDMR experiments are generally based on top illumination and ohmic contacts close to the focal spot \cite{bourgeois2020photoelectric}. We explore an alternative illumination scheme in which the laser beam enters laterally through the side of the diamond sample just below the surface, forming an optical path between the two electrodes (see Method, Setup). In Figure \ref{exp1}-a, we plot the current-voltage (I-U) characteristics for different laser powers. All curves start with a first, almost linear, increase as a function of voltage, followed by a second, less steep one, after an inflection point that increases as laser power is raised from 100 mW to 400 mW. For a laser power of 400 mW, the inflection point occurs at a voltage of 100 V. Given the 200 \textmu m gap between the electrodes, this voltage corresponds to an average electric field between the electrodes of less than half the value at which the carrier velocity saturates in diamond \cite{wort2008diamond}, suggesting Schottky-like behavior for our contacts. Under this assumption, we model our metal-semiconductor-metal system as two back-to-back Schottky diodes, one of them being forward-biased and behaving like an ohmic contact, the other one being reverse-biased and ruling the overall PC (Figure \ref{exp1}-a).

In order to identify the conduction state of the two contacts, we apply a magnetic field along one of the four $\langle 111 \rangle$-directions of the diamond lattice with a magnetic field gradient to create a correspondence between the NV$^-$ resonance frequency and the spatial position in the field of view \cite{magaletti2022quantum}. The magnetic field gradient was tuned so that the $|0\rangle \rightarrow |-1\rangle$ spin transition of the NV$^-$ center family aligned with the magnetic field was at 1.98 GHz for the area corresponding to electrode A (Figure \ref{exp1}-b) and at 2.02 GHz for the area corresponding to electrode B (Figure \ref{exp1}-c). When +150V is applied to electrode A, the PDMR spectrum exhibits a resonance at 1.98 GHz, similar to the ODMR spectrum for the NV$^-$ centers under electrode A (Figure \ref{exp1}-b), showing that contact A drives the current through the structure and is thus reverse-biased while contact B is forward-biased. Similarly, when the voltage is switched to +150 V on electrode B, the PDMR spectrum shifts to a resonance at 2.02 GHz, identical to the ODMR spectrum obtained for the NV$^-$ centers below electrode B (Figure \ref{exp1}-c), showing that contact B is now reverse-biased, while A is now forward-biased.

The fact that a positive voltage induces a reverse-bias on the Schottky contact indicates that the diamond sample behaves as a p-type semiconductor under green illumination \cite{sze2021physics}. In our sample, the dominant dopant is nitrogen, with residual traces of boron. It has been shown that in the case of a diamond sample where nitrogen overcompensates for boron (as in the case of parasitic incorporation), a p-type semiconductor can be obtained under UV illumination \cite{grunwald2021photoconductive}. We adapted the model of Grünwald et al. \cite{grunwald2021photoconductive}, to account for a different charge generation mechanism: electron-hole pairs generated by NV centers switching from their negatively charged state to their neutrally charged state under green illumination (see Supplementary Information (SI), section 1). Due to a faster electron recombination rate in the presence of nitrogen-overcompensated diamond material, a higher concentration of holes is required to satisfy electroneutrality, resulting in a p-type semiconductor under green illumination.

\begin{figure*}[h]
    \centering
    \includegraphics[width=\textwidth]{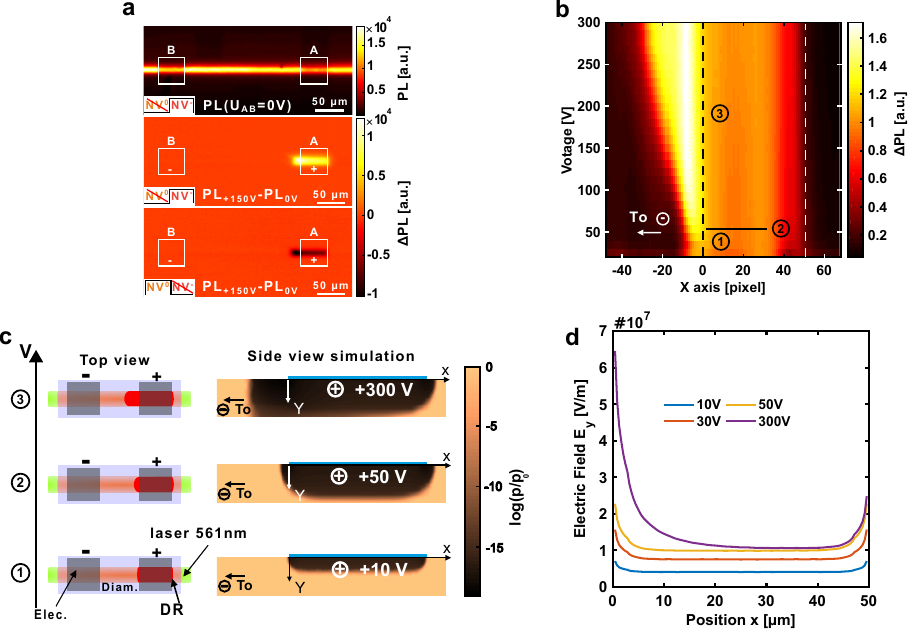}
    \caption{\textbf{Visualization of DR under the positive electrode.} \textbf{a}, Observation of the DR using NV${^-}$ and NV${^0}$ optical filter. Top: PL reference image at 0V using a NV$^{-}$ bandpass filter. Middle: $\mathrm{\Delta PL}$ image obtained by subtracting the PL reference image at 0V to the PL image obtained when applying +150V on electrode A using a NV$^{-}$ bandpass filter. Bottom: $\mathrm{\Delta PL}$ image obtained for the similar procedure but using a NV$^{0}$ instead of a NV$^{-}$ bandpass filter (see Methods, “DR-optical imaging”). \textbf{b}, Normalized $\mathrm{\Delta PL}$ profile at the positive electrode with the bias voltage showing the DR-lateral extension (see Methods, “DR-optical imaging”).  \textbf{c}, Scheme (top view) and simulation (side view) of the different stages of DR extension as bias voltage increases. \textbf{d}, Electric field distribution close to the positive electrode for some values of bias voltage.} 
    \label{exp2}
\end{figure*}

This p-type conductivity under illumination is reflected at the graphitic-diamond junction by the stabilization of the NV centers in their negative charge state, particularly when a bias voltage is applied. Figure \ref{exp2}-a shows the PL profile signal when the applied voltage is turned on or off. A higher PL level when using a NV$^{-}$ filter and a lower PL level with a NV$^{0}$ filter are observed under the positive electrode in figure \ref{exp2}-a when turning on the voltage. These phenomena can be explained by the band bending, which stabilizes the NV negative charge state in the DR, assuming our diamond is p-type under illumination (see SI, section 2) \cite{schreyvogel2015active}.

The extension of the DR as a function of the applied voltage for an illumination power of 100 mW, is shown in Figure \ref{exp2}-b. First, at a bias voltage below 50 V, the lateral extension of the DR towards the negative electrode remains almost unchanged. Second, at higher bias voltages, the DR extends laterally towards the negative electrode with an expected square root dependence on the voltage (see SI, section 3) \cite{sze2021physics}. To better understand this behavior and obtain the vertical and lateral extension profile of the DR with the bias voltage, we used COMSOL Multiphysics simulations assuming that laser illumination below the electrodes results in a homogeneous hole density $p_0 = 3.5 \times 10^{14}$ cm$^{-3}$ over a depth of 10 µm below the diamond surface, according to the beam size (see SI, section 4, Figure S4-a). As seen in Figure \ref{exp2}-c, at low bias voltage, the DR first extends mainly vertically (i.e., perpendicularly to the electrode), which corresponds to stage 1 in Figure \ref{exp2}-b,c. At a certain voltage close to 50 V, referred to as stage 2 in Figure \ref{exp2}-b,c, the DR reaches the bottom limit of the illuminated part. As the area of the diamond that is not illuminated behaves like an insulator, the DR can no longer continue to extend perpendicular to the crystal surface. Finally, increasing the bias voltage further leads to a lateral extension of the DR toward the negative pole (stage 3, Figure \ref{exp2}-b,c). These different regimes of the DR as a function of bias voltage also have an important effect on the electric field at the electrode surface. As shown in Figure \ref{exp2}-d, the electric field profile is, as expected, almost uniform across the entire electrode, except at the left and right edges where an edge effect occurs. By increasing the reverse bias voltage, the electric field away from the edges increases until the DR reaches the bottom limit of the illuminated part, after which it remains constant if the bias voltage is further increased. The maximum electric field under the positive electrode, far from the edge, is a direct function of the hole concentration. For the same reason, this electric field is also sensitive to the NV$^-$ center spin resonances, as these affect the hole concentration (see SI, section 4, Figure S5-a). This mechanism plays an important role in the PDMR signal obtained from such back-to-back Schottky structure.

\section{The PDMR contrast using Schottky contact}

Upon illumination, the graphitic-diamond-graphitic structure behaves like two metal-semiconductor (M-SC)-p Schottky junctions in back-to-back configuration. Consequently, the overall current will be set by the reverse current allowed at the reverse-biased contact. The reverse current in a Schottky diode is mainly due to thermionic emission, which is given by \cite{sze2021physics}:

\begin{equation}
    I_{th} = I_0 e^{\frac{-q (\phi_1- \Delta\phi)}{kT}} (1-e^{\frac{-q U_1}{\eta kT}}),
    \label{inverse}
\end{equation}
\noindent where $q$ is the elementary charge constant, $T$ is the absolute temperature, $k$ is the Boltzmann constant, $U_1$ is the voltage drop at the reverse junction (where the DR is located), so  $U_1\approx U$ is the applied voltage (see in SI, section 3). Here, $\phi_{1}$ represents the barrier height of the reverse-biased Schottky junction, and $\eta$ is the ideality factor. $I_0$ relates to the Richardson constant $A^*$ for holes, the absolute temperature $T$, and the effective surface of the contact $A$: 
\begin{equation}
    I_0 = AA^* T^2,
    \label{Rich}
\end{equation}

\noindent In addition, an image force lowering (IFL) $\Delta \phi$, reduces the barrier in the presence of an electric field, thus affecting the charge carrier emission through the barrier \cite{sze2021physics}. $\Delta\phi$ reads as follows: 
\begin{equation}
    \Delta \phi = \sqrt{\frac{qE}{4\pi \epsilon_0 \epsilon_s}}.
    \label{image-force}
\end{equation}
\noindent where $E$ is the electric field at the M-SC interface part \cite{sze2021physics}, $\epsilon_0$ is the dielectric constant of vacuum, and $\epsilon_s$ is the diamond relative permittivity.

\begin{figure*}[h]
    \centering
    \includegraphics[width=0.85\textwidth]{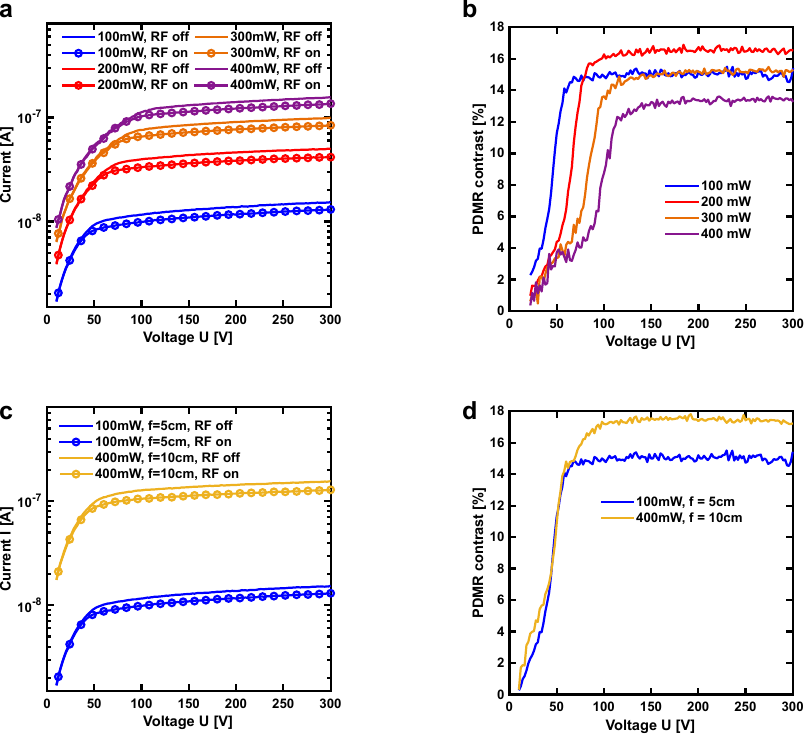}
    \caption{\textbf{The evolution of PDMR contrast.} \textbf{a}, I-U characteristics for some values of laser power in the presence and absence of RF at 2.87 GHz in semi-log scale. \textbf{b}, Evolution of PDMR contrast corresponding to \textbf{a}. \textbf{c}, Comparison in semi-log I-U characteristics with the same illumination intensity: 100 mW with a 5 cm focal lens and 400 mW with a 10 cm focal lens. \textbf{d}, Evolution of PDMR contrast corresponding to \textbf{c}.}
    \label{exp3}
\end{figure*}


Figure \ref{exp3}-a shows the I-U characteristics at resonance (2.87 GHz under zero magnetic field) and out of resonance for different laser illumination powers. For a given illumination power, the application of a radiofrequency (RF) field in resonance with the NV$^-$ centers reduces the current and shifts the inflection point to a lower voltage value. The change of inflection point at resonance results in three different regimes in PDMR contrast, which are particularly visible for the 400 mW power in Figure \ref{exp3}-b, and which are well described by the model in SI (section 4, Figure S5). This behavior can be understood considering the concentration of holes in the illuminated part, which is related to the spin-dependent modulation of the electric field at the reverse-biased contact (see SI, section 4). Indeed, applying a resonant RF excitation to the NV$^-$ centers decreases the concentration of holes produced in the illuminated part. This reduction in hole concentration leads to an extension of the DR, then reducing the electric field at the reverse-biased contact. Increasing the bias voltage further accentuates the electric field difference in the presence or absence of a resonant RF excitation, thus increasing the PDMR contrast, until the DR reaches the lower limit of the illuminated portion below the electrode where, as seen previously in Figure \ref{exp2}-d, the electric field can no longer increase. Beyond this point (e.g. 100 V at 400 mW illumination power and 25 dBm MW power at 2.87 GHz, Figure \ref{exp3}-b), the PDMR contrast increases rapidly as the electric field continues to increase with bias voltage only in the absence of resonant RF excitation until the DR finally reaches the lower limit of the illuminated area (e.g. 125 V at 400 mW illumination power, Figure \ref{exp3}-b). For this last regime, the electric field difference (between the non-resonant and resonant situations) and the associated PDMR contrast remain almost constant. The change observed in reverse current as a function of bias voltage after the inflection points is attributed to edge effects on the electrode, where the electric field continues to increase when the DR extends longitudinally and no longer vertically, as previously observed in Figure \ref{exp2}-d. The PDMR contrast evolution profile with bias voltage is very similar to that obtained by using the electric field from simulations to extrapolate the change in thermionic emission induced by NV$^-$ spin resonances at 2.87 GHz, assuming that optical pumping produces a 10 µm hole-doped volume below and between the electrodes, and assuming that RF resonant excitation at 2.87 GHz and 25 dBm decreases the hole concentration in the illuminated volume (see SI, section 4, Figure S5-a).

We then highlight the impact of the thickness of the hole layer produced by the NV centers upon illumination. We compare the off- and on-resonance I-U characteristics (Figure \ref{exp3}-c) as well as the PDMR contrast as a function of bias voltage for two different values of the beam size and the same laser intensity (Figure \ref{exp3}-d). Using two lenses with different focal lengths (5 cm and 10 cm, resulting in a twofold variation in beam waist) while we adjuste the laser illumination power (100 mW and 400 mW) to keep constant the laser intensity and thus the hole concentration produced upon illumination. As expected from the simulation (see SI, section 4, Figure S6), the inflection point is shifted towards higher voltages as the beam size increases (Figure \ref{exp3}-c), because the DR has more room to extend vertically under the positive electrode. For the same reason, a larger beam size enables a greater electric field difference in the saturation regime, enhancing PDMR contrast. This highlights that the beam size, particularly in side illumination experiments, can serve as an additional degree of freedom to optimize the electrical readout of NV$^-$ ensemble spin resonances.

\section{Conclusion}\label{sec12}

Beside the ohmic-contact-based electrical readout, which monitors changes in the spin-dependent photoconductivity, our two-Schottky-contact approach utilizes the same spin-dependent process but with a fundamentally different current mechanism. We have seen that it is possible to locally transform an insulating diamond crystal lightly doped in B and N into a p-type semiconductor material by exploiting both the dual donor/acceptor nature of the NV center and the charge recombination mechanisms in this type of crystal. Combined with the space charge effects arising from the Schottky nature of the graphitic-diamond contacts, the result is a thermionic current that is ruled by the spin resonances of the NV$^-$ centers. Compared to the ohmic approach, electrical readout based on this back-to-back Schottky structure presents specific features. It allows to obtain local information on the NV$^-$ spin resonances because the electric field governing the thermionic emission is localized in the depletion region. This local dependence makes optical addressing possible using an illumination scheme similar to that used in this work or, by illuminating through other types of semi-transparent electrodes, producing a similar Schottky contact. Such a structure could considerably facilitate wiring, for example by using only two large parallel plates instead of an array of electrodes. In addition, one could also consider adapting the electrical readout scheme slightly for something more specific to the Schottky contact by exploiting the change in capacitance induced by the modulation of depletion region size induced by NV$^-$ spin resonances instead of thermionic current modulation. Finally, as current emission through a Schottky barrier is a highly nonlinear field-dependent process, it is also possible to exploit the field effect induced either by the illumination scheme or by the electrode geometry to optimize PDMR contrast. Our results shed new light on the close link between the optical and electrical properties of NV-doped diamond upon illumination, highlighting the potential of a new type of NV-based quantum sensor relying on electrical readout from Schottky contacts.
\section*{Methods} 
\subsection*{Setup}

We implement a setup for simultaneous ODMR imaging and PDMR monitoring. The diamond crystal used in this experiment is an optical grade diamond crystal from Element Six (E6) containing 1 ppm of nitrogen, a few ppb of NV (corresponding to an approximate creation yield of around 0.3\% \cite{edmonds2012production}), and around 50 ppb of boron which results from parasitic doping during the CVD growth process. $\langle 100 \rangle$ and $\langle 110 \rangle$ side facets are optically polished. The NV centers are pumped in a continuous wave regime by a 561 nm laser (Cobolt Jive). The laser wavelength is chosen to avoid the repumping from the NV$^-$ metastable state to the conduction band \cite{hruby2022magnetic}. The PL is collected by a microscope objective (Mitutoyo P Plan Apo 10X), filtered by a bandpass NV$^0$ filter (Semrock, AF01-584/40) or NV$^-$ filter (Semrock, FF01-698/70), and detected by a CMOS camera (IDS Ueye). In order to measure the PC, we realize on the diamond crystal two graphitic electrodes measuring 50$\times$50 µm$^2$, separated by 200 \textmu m. The graphitic phase was obtained by transforming diamond to amorphous carbon by implanting Xenon with a focused ion beam \cite{konov2012laser, villaret2023efficient}. A Xenon dose of 4 $\times$ 10$^{15}$ cm$^{-2}$ with an acceleration energy of 30 keV produces electrodes with a thickness of a few nanometers, thus leaving the area transparent enough to realize ODMR imaging of the NV$^-$ centers beneath the electrodes. The laser enters and leaves the sample by the side facets, bridging the two electrodes (Figure \ref{energyband}-b). The PC is measured by a high voltage source measure unit (SMU) (Keithley, 2470). The RF field is brought into proximity of the NV centers by a loop antenna centered with respect to the two electrodes (see Figure \ref{energyband}-b). A neodymium permanent magnet generates a static magnetic field to lift the degeneracy between the four NV center families in the diamond crystal.

The (561 nm) laser beam used in this study has an initial diameter of 700 \textmu m. The beam passes through a beam expander (magnification: 5X) and is then focused onto the diamond by means of a lens. Due to the Gaussian profile of the beam, using a 5 cm lens produces a focal spot with a diameter of around 10 \textmu m.

\subsection*{DR-optical imaging}

Charge state stabilization of the NV$^-$ centers due to band bending can be exploited to optically reveal the DR profile that results of the graphitic-diamond Schottky contact under illumination using a NV optical filter and subtracting the images when the bias voltage is turned from on to off. In our case, applying an external bias voltage leads to the stabilization of the illuminated NV centers in their negative charge stage; thus using a NV$^-$ (respectively NV$^0$) bandpass filter produces a positive (respectively negative) local change in photoluminescence (denoted $\mathrm{\Delta PL}$) when a bias voltage is applied, which corresponds to the region where the band bending occurs (i.e. the DR) (Figure \ref{exp2}-a).

\subsection*{Evolution of the depletion region }

We exploit the $\Delta$PL image using a NV${^-}$ filter for highlighting the DR extension with the bias voltage using a very simple data processing. For each bias voltage, we extract a $\Delta$PL profile along the illumination path between the electrodes that corresponds to the x-axis, averaging vertically the $\Delta$PL signal over five pixels and taking for the origin of the x-axis the left edge of the positive electrode facing the negative electrode. The $\Delta$PL profile was then normalized using the $\Delta$PL value obtained at the middle of the positive electrode as an arbitrary reference (x=25 in Figure \ref{exp2}-b). We use this procedure to produce the data in Figure \ref{exp2}-b.

\subsection*{ODMR and PDMR contrasts}

The PDMR contrast ($C_{PDMR}$) is given by:
\begin{equation}
  C_{PDMR} = \frac{I_{OFF} - I_{ON}}{I_{OFF}},
    \label{def_contrast_PDMR}
\end{equation}
\noindent where $I_{ON/OFF}$ is the photocurrent measured when a RF field is respectively on resonance and out of resonance with the NV$^-$ centers. Similarly, the ODMR contrast is defined as:

\begin{equation}
  C_{ODMR} = \frac{PL_{OFF} - PL_{ON}}{PL_{OFF}},
    \label{def_contrast_ODMR}
\end{equation}
\noindent where $PL_{ON/OFF}$ is the photoluminescence measured when a RF field is respectively on resonance and out of resonance with the NV$^-$ centers.

\section*{Data availability} 
The data supporting the findings of this study can be obtained from the corresponding author upon reasonable request.
\section*{Acknowledgements} 

This project has received funding from the European Union’s Horizon Europe research and innovation program under grant agreement No. 101080136 (AMADEUS), the EIC Pathfinder 2021 program under grant agreement No. 101046911 (QuMicro), and the EMPIR project 23NRM04 (NoQTeS). We acknowledge the support of the Agence Innovation Défense under grant agreement ANR-22-ASM2-0003 (SPECTRAL). 

The authors acknowledge A. Rowe, J. Achard, J. Barjon, D. Eon, J. Pernot and H.-J. Drouhin for fruitful discussions.


\section*{Conflict of Interest}
The authors have no conflicts to disclose.
\section*{Author contributions} 
X.L., L.M., S.M., M.S., J.-F.R. and T.D. contributed to the design and implementation of the research, to the analysis of the
results and to the writing of the manuscript.

\section*{Additional information}     
\textbf{Supplementary information} The online version contains supplementary material available at ...

\bibliography{sn-article}

\end{document}